# Multiple Uncertainties in Time-Variant Cosmological Particle Data

Steve Haroz[*]     Kwan-Liu Ma[†]
*University of California at Davis*

Katrin Heitmann[‡]
*Los Alamos National Laboratory*

**Abstract**

Though the mediums for visualization are limited, the potential dimensions of a dataset are not. In many areas of scientific study, understanding the correlations between those dimensions and their uncertainties is pivotal to mining useful information from a dataset. Obtaining this insight can necessitate visualizing the many relationships among temporal, spatial, and other dimensionalities of data and its uncertainties. We utilize multiple views for interactive dataset exploration and selection of important features, and we apply those techniques to the unique challenges of cosmological particle datasets. We show how interactivity and incorporation of multiple visualization techniques help overcome the problem of limited visualization dimensions and allow many types of uncertainty to be seen in correlation with other variables.

Animations as well as additional information are available at http://vis.cs.ucdavis.edu/~haroz/cosmology

**KEYWORDS**: Visualization applications, cosmology, uncertainty visualization, parallel coordinates.

**INDEX TERMS**: H.5 [Information Interfaces and Presentation]; I.3.8 [Computer Graphics]: Miscellaneous – Time-Variant Vertices

## 1 INTRODUCTION

The visual system is a powerful tool for pattern recognition, yet it requires its input to be particularly formatted. Spatial perception is one of our more capable analysis tools, yet highly detailed spatial data imposes a limitation on the flexibility that often characterizes some of the more powerful information visualization techniques. Given the complexities already present in the visualization of spatial data, presenting multiple types of uncertainty along with spatial information can accordingly provide a unique challenge.

In this paper we explain how uncertainty, when quantifiable, can be applied as additional dimensions. As visualization has been established as an effective technique for data mining [1], we in turn advance these visualization approaches to find new and insightful information about a dataset that has multiple sources of uncertainty. Furthermore, we explore the sparsely studied area of uncertainty over time.

The driving force behind this study is a series of cosmological particle datasets with multiple dimensions of data as well as time variance and uncertainty across all of the dimensions. We specifi-cally examine past scenarios wherein a particle visualization requires a defined spatial element and multiple additional dimensions [2], [3]. We then explain how multidimensional visualization techniques can apply to the uncertainty of such data. These techniques can help a user examine correlation patterns between dimensions and use the insight gained to interactively select a time step and dimension for emphasis and detailed exploration. The benefit of this approach is that scientists can visualize multiple types of uncertainty and can mine each in the same manner that they examine other types of data. Furthermore, they can visualize uncertainty without limiting their ability to visualize other variables of interest. We demonstrate how the multiple types of uncertainty in the spatial and temporal dimensions of these cosmology datasets can be visualized and explored with the ultimate goal of helping scientists answer their own questions.

## 2 UNCERTAINTY VISUALIZATION

The level of data uncertainty can be a crucial component in making an informed decision. If the goal of visualization is to provide insight into data, then the certainty of that data should also be presented. From MRIs used by medical professionals to wind speeds used by fire-fighters, misconstrued confidence or simply a lack of consideration of data uncertainty can have life threatening implications. Nevertheless, the presentation of information certainty (or the lack thereof) can be difficult when the mere problem of effectively presenting absolute data alone necessitates a complex solution. Whether ignoring uncertainty or presenting it while sacrificing another data variable, the consequences of incomplete information may be unacceptable for the intended viewer. The field of uncertainty visualization has made great strides recently [4], and with ever larger datasets being made available, new applications of those techniques are becoming an ever more feasible body of research.

A myriad of causes can result in uncertainty. Down-sampling, inconsistent measurement quality, and variations in simulation calculations can all affect data precision [4], [5], [6]. Visualizing data while acknowledging its ambiguity is a difficult feat, as the representation of the uncertainty can easily interfere with the representation of the data itself. Nevertheless, rendering uncertainty in combination with accurate 3D data is particularly necessary for many scientific visualizations. In turn, visualizing uncertainty in a bound 3D space can present a difficult challenge, as displaying the ambiguity can use up important features of volume visualization such as color or opacity. Workarounds for this problem can include using texture, blurred rendering, or noise; however, such approaches would be difficult to apply to data that is already noisy or that has detail greater than the resolution of the screen.

Rendering the uncertainty of a shape or surface presents a similar problem to volume uncertainty, as the uncertainty display can occlude or interfere with the spatially constrained data. Texture, color, bump mapping, and point clouds are possible resolutions to the problem [5], [7]. Bump mapping a surface can be particularly effective at portraying some types of surface uncertainty; it intuitively adds noise and ambiguity to the surface. Though effective for some datasets, bump mapping provides little information about the type of uncertainty. Scenarios in which the statistical information or source of the uncertainty is of major interest may not benefit from this approach. The effect could also be confused with the surface itself. Typical point clouds around a surface can have a similar problem. Though their noisiness can differentiate them from a surface, the static nature of the visualization makes interpreting them as an actual noisy surface entirely plausible. Ultimately, all of these methods have great merit for visualizing particular types of uncertainty; choosing one must depend on the type of additional data that the designer wishes to visualize.

[*]sharoz@ucdavis.edu
[†]ma@cs.ucdavis.edu
[‡]heitmann@lanl.gov

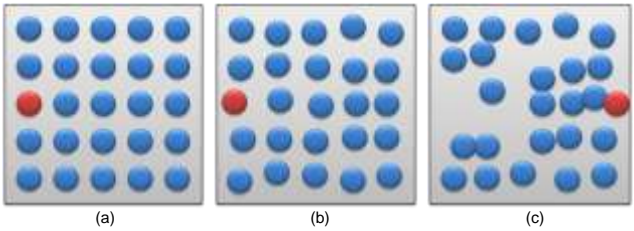

Figure 1. (a) The particles are arranged evenly on a 3D grid. (b) They are then perturbed slightly to satisfy the cosmological initial conditions. (c) The simulation then moves the particles based on gravitational forces forming dense clusters and sparse regions. Notice that that the red particle moves so far left that it wraps around the edge. In reality, the particles are of course much smaller and the simulation has no collisions, i.e. particle scattering is very small and ideally does not occur.

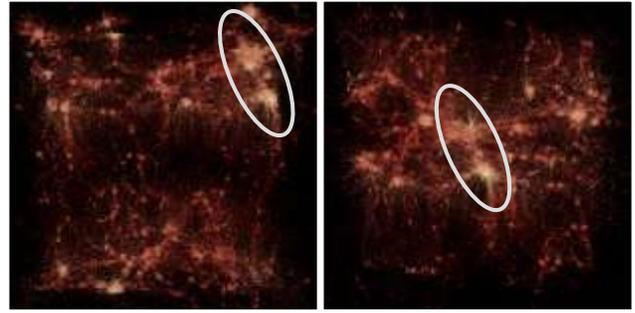

Figure 2. These images show the FLASH dataset. Each image shows a different wrapping offset of the same data from the same view, and the circled clusters are the same in each image. The clusters and sparse regions of the particles are clearly visible. The coloring is based on velocity magnitude which correlates highly with dense clusters of particles. Red represents low velocity, while light yellow represents high velocity.

With the variety of sources of uncertainty in data, the type of uncertainty should be an important consideration in visualization design. Olston and Mackinlay explored the differences between uncertainty defined statistically (for example via a mean and standard deviation) and uncertainty defined as a range between two values [8]. They explain that the standard statistically oriented method of error bars is unsuited for displaying ranged uncertainty. In its place, they propose using a technique that they call "ambiguation" to present bound uncertainty. Using simple 2D graphs, they demonstrate their method of clearly defining the minimum value and graying the area between the minimum and maximum. They also provide examples of bounded and unbounded uncertainty in graphs that have a fixed area (such as a pie chart). In these more rigid cases, the uncertainty lies in the boundary between two regions. The suggested approach is to make a blurred region or shade the borders based on the level of uncertainty. The bounded range of star positions and constellations has recently been visualized by Li et al. [9].

## 3 COSMOLOGICAL SIMULATION COMPARISON

As an example 3D dataset with significant uncertainty, we examine the results of a cosmological particle simulation generated as part of a study of cosmological simulation robustness [10]. The goal of the dataset is to examine the inconsistencies between different simulations that begin with the same initial conditions [2]. The simulation begins as $256^3$ particles arranged evenly on a cubical grid. The particles are then moved by small amounts to establish the correct cosmological initial conditions (see figure 1), which are constrained by observations of the cosmic microwave background and the distribution of galaxies on large scales. The particles then move under the influence of gravity in an expanding universe for a number of time steps until the current epoch is reached.

As is typical for cosmological simulations, periodic boundary conditions are imposed. When a particle moves past one edge of an axis, it appears on the other end. This property is intriguing from a visualization perspective because the axes wrap. As we will discuss later in the paper, we address this important hurdle in our visualization implementation.

The dataset also includes important elements of uncertainty that have not yet been visualized. The sources of uncertainty are the numerous simulator codes used to compute the inter-particle forces and the approximated values that drive them. Some simulators make use of hierarchical sampling of the system phase space distribution function [11], [12], while others simplify by distance [11], [13]. Each simulation was run separately, and as a result of their implementation differences, every simulation produced slight deviations in the particle distributions and final velocities. Past comparisons of the codes have found quantitative differences in the simulation results [10], [11].

Past approaches have made effective progress in visualizing similar types of particle data. These methods have merit in their effective portrayal of the particles position and velocity [2]. However, we now have the benefit of hindsight and newer technology. Earlier approaches to visualize even a single dataset tended to rely on hierarchical groupings of the data, which required preprocessing and only provide complete detail at the lowest levels [14].

Extending the single-dataset efforts, Ahrens et al. studied the visualization of multiple datasets via a side-by-side comparative analysis [2]. They visualized two individual datasets in separate views with linked transforms. Each dataset was generated by a different simulation code, but the initial conditions were the same. Although this method can be useful if proceeded by a focused serial search through the images, it unfortunately relies upon the assumption that we can proficiently detect differences in side-by-side images. Testing a principle known as *change blindness*, perception researchers have found that we are surprisingly poor at detecting changes between images if a small interruption exists between them [15]. That interruption can be as insignificant as the brief, split-second blindness that occurs during a saccadic eye movement. In other words, significant changes can go unnoticed when shifting focus from one image to another. The optimal presentation of difference would lack even the slightest spatial and temporal interruptions.

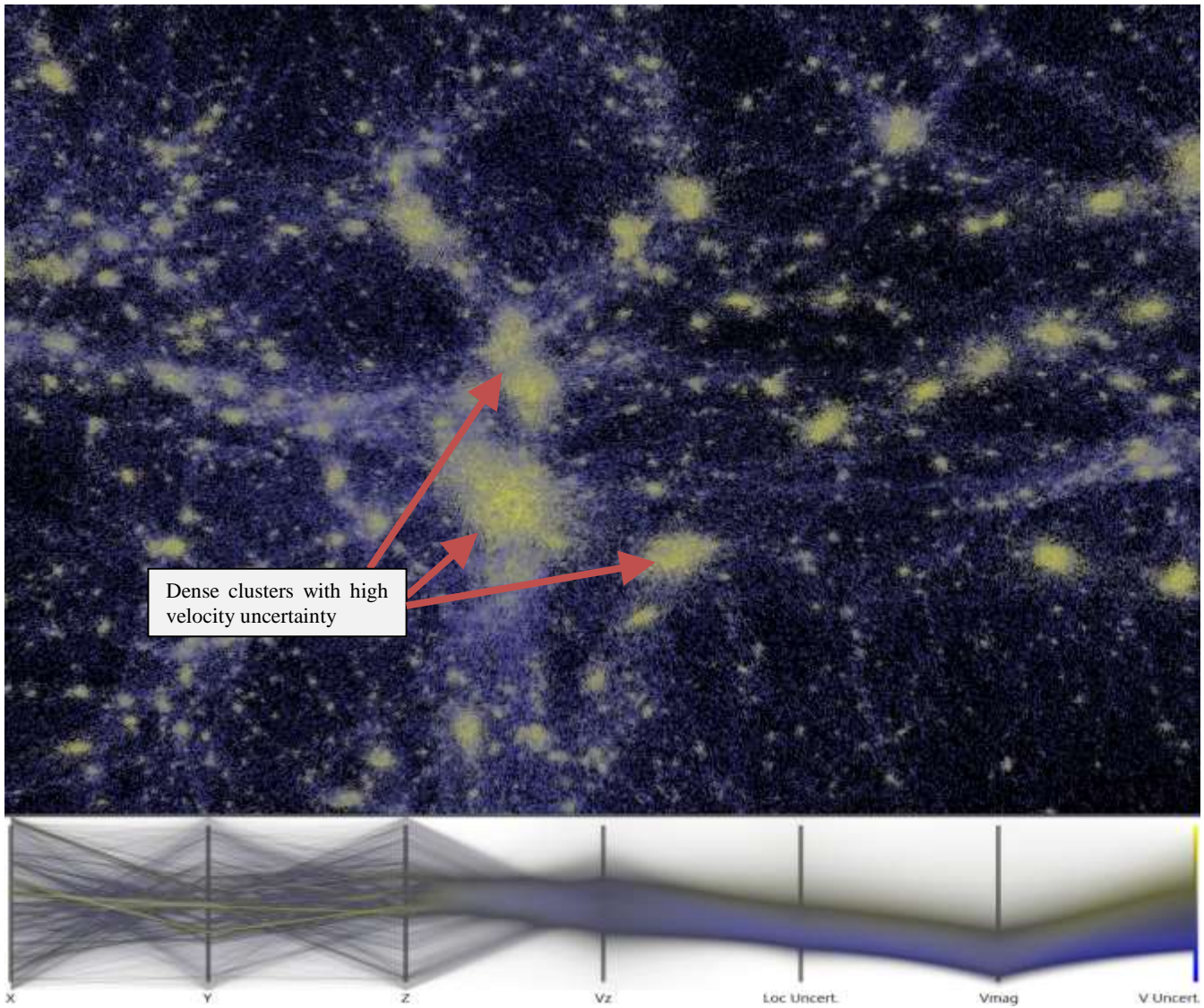

Figure 3. This zoomed view shows a comparison of the FLASH and HOT simulation results with the particles colored based on the velocity uncertainty. Yellow particles have the highest velocity uncertainty, while blue particles are more certain. The parallel coordinates view on the bottom shows a smooth gradient along the leftmost axes, velocity magnitude (Vmag) and location uncertainty (Loc Uncert). The bright yellow lines along the center of X, Y, and Z axes (on the left) represent the position of the dense and highly uncertain clusters in the center of the view. The difference in the force resolution of the two simulators results in high variation in the dense regions, which are bright yellow.

### 3.1 Dataset and Visualization Interface

Each particle in the dataset has a position and velocity in 3D space. The particles also have an ID number, so a single particle can be identified in multiple simulation results.

The particles have moved under the force of gravity and formed large scale structures. The overall structure has three main elements: highly dense regions – the so-called clusters, highly sparse regions – the so-called voids, and filaments. The general features of the large scale structures (e.g. the positions of clusters) and the shapes of voids are the same in all the different simulations, while the exact particle positions and densities can vary. Within the dense clusters, particles with varied velocity can be in extremely close proximity, so voxelizing and rendering this data as a volume is not desirable.

For one view, our implementation renders each particle as a small line with a length and orientation that correspond to the particle's velocity. The middle of that line is the location of the particle. The location property, however, cannot be naively taken at face value. The wrapping of the axes means that important information such as clusters may be split by the edges of the bounding cube. To address this potential problem, we created a set of three sliders. Each corresponds to one of the axes and represents an offset used to wrap the axis. The user can adjust the sliders to move the interesting information away from the edges (see figure 2).

### 3.2 Multivariate Representation

Next to the primary spatial visualization we include a parallel-coordinates view of the data that also acts as a control mechanism. This view allows numerous dimensions of the data to be simulta-neously visualized in reference to each other [16]. The axes include the positional dimensions, velocity components, velocity magnitude, positional uncertainty, and velocity uncertainty (see the bottom of figure 3 and figure 4). The user can select any of the dimensions as the coloring basis for both views.

### 3.3 ADDING THE UNCERTAINTY DIMENSION

One means of making an object appear uncertain is to blur the representation. This technique can be applied to bound uncertainty (finite ranged) or unbound uncertainty (normally distributed) [8]. However, the sheer quantity of particles makes their average screen representation smaller than a pixel on even the largest displays. Blurring would be unnoticeable due to the existing aliasing and lack of shape, or it would create too much occlusion. Blurring and altering glyph shape [17] are both methods that typically represent uncertainty, but they are simply not applicable to this type of dataset. We therefore make use of the classic visualization property, color, to represent uncertainty, which figure 3 presents.

Binding color to uncertainty would make that powerful visualization feature unavailable to other dimensions. Past researchers have effectively associated color with velocity magnitude in this dataset and found insightful information such as a strong correlation with density [2], [11]. To unconstrain color, we allow the users to interactively select a parallel coordinates dimension. This ability allows them to dynamically choose which feature is important, so they can focus on the information that they want.

When a new dimension is selected for coloring, both the spatial view as well as the parallel coordinates view reflect that change. The particles in the spatial view are colored based on their associated value, and the parallel coordinates view is also redrawn under the new coloring scheme. The colored parallel coordinates allow the user to find correlations and patterns between dimensions that might be difficult to see using only a colored dimension in a spatial view. Using parallel coordinates as a selection tool and interface for a spatial view has been shown to be particularly effective for particle visualizations [3]. To make the parallel coordinates accurately render over sixteen million points, we used a 32-bit-per-channel frame buffer that allows for very low alpha values. In turn, the lines blend to form a smooth anti-aliased display.

To add the uncertainties as dimensions, we need to quantify them. The user should be able to visualize, explore, and interact with uncertainty in both position and velocity in the context of all of the data variables. For each particle, the application averages the vectors of each value and finds the standard deviation. It then simply calculates magnitude of the resulting standard deviation vector. If only two datasets are loaded, the standard deviation is approximated as the difference. While calculating the velocity uncertainty is straight forward, calculating the positional uncertainty requires taking the axial wrapping into account. A particle whose positions slightly straddle the edge of an axis is appropriately calculated as having a small uncertainty.

### 3.4 FINDINGS

As an initial test of the visualization system, we confirmed previous findings [2] that the particles with high velocities clustered in the very dense regions as is apparent in figure 3.

When the velocity uncertainty, the position uncertainty, or the velocity magnitude is selected as the colored axis, a smooth vertical gradient appears between those axes (see the bottom right of figure 3). This pattern shows that the values are highly correlated [16]. The result is reasonable to expect, as a high velocity would create a large positional change resulting from only a slight variation in position. We also tried normalizing the velocity uncertainty by calculating it based on velocity orientation alone (e.g. without incorporating velocity magnitude). The difference was minimal, and the results showed that density, velocity magnitude, and positional uncertainty also correlate strongly with uncertainty in velocity orientation.

A correlation exists between position along any arbitrary axis and the velocity along that axis. The particles are moving toward a group of highly dense clusters. As can be seen in figure 4, coloring the particles based on angular distance from a particular vector, such as the Z axis, results in a somewhat sharp color border on either side of the cluster. Such insight would be difficult to find if color was locked to vector magnitude.

## 4 VARIANCE OVER TIME

Another aspect of the simulations that the scientists wish to explore is the impact of different initializations of the particles. Scientists want to follow the evolution of the universe from the very first moments to today, yet the accuracy of the simulators places limitations on the starting time of the simulation. In the early time steps of the simulation, the small perturbations can be too miniscule to be accurately stored by the single precision floats used by the simulation codes. The scientists therefore need to start the simulation a few million years after the Big Bang (the universe is roughly 13.66 billion years old). Fortunately, the physics during that initial period is well understood (due to its linearity). They can use the Zel'dovich approximation [18] to skip over the initial time steps with the added bonus of reducing simulation time. The caveat of the approximation is that using it at a time too close to the current epoch can adversely impact the accuracy of the simulation, as structures do not seem to form in the same way. Furthermore, the accuracy of the Zel'dovich approximation requires that particle paths do not cross. The questions that scientists want to examine are how long the approximation should last and how the approximation impacts the latter time steps of the simulation.

The physical expansion of the universe is scaled out of the simulation by the scale factor of the expansion, a. The simulation box therefore always contains a constant "comoving" volume even though the actual physical volume is always increasing. Each time step represents a particular epoch in the Universe's history defined by the particular value of the scale factor or alternatively the redshift, $z=1/(1+a)$. This dataset has multiple series of time-variant data. Beginning with the initial grid, each time series extends the Zel'dovich approximation to a different time step. Each series then runs through the simulation with several time steps being outputted along the way (see figure 5). The result is a dataset with the almost unstudied property of time-variant particle uncertainty. Each time step has position and velocity uncertainty that vary according to approximation length.

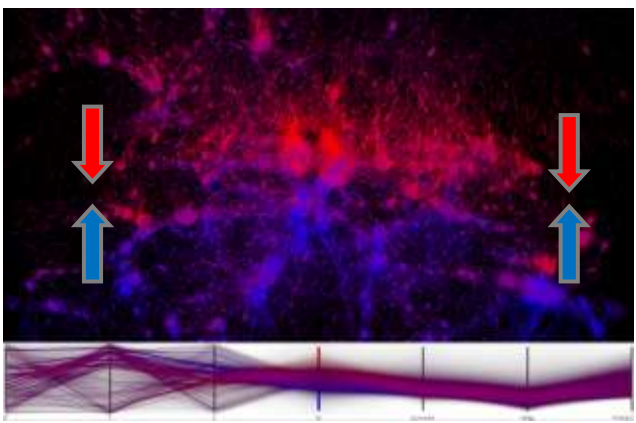

Figure 4. The particles are colored based on the projection of their velocities along the Z axis. The blue particles are moving up, and the red particles are moving down. The image shows that the particles are gravitating toward the dense regions in the center. The positional axes (left three) show a clear correlation due to the bands of different color, whereas the uncertainty and velocity magnitude axes (right three) have a consistent intermediate color.

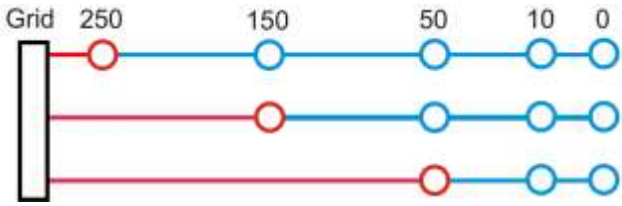

Figure 5. This diagram illustrates the correlation of datasets. The grid (black box on the left) is the unperturbed locations for all of the particles. The Zel'dovich approximation (red lines) progresses the particles to a particular time step. The simulation (blue lines) then progresses until the present time (time 0). Each circle represents a dataset that we have. We interpolated to obtain intermediate time steps.

### 4.1 Hardware Accelerated Implementation

For scientists to explore this problem, they need to be able to interactively examine how the data changes over time and view the variation caused by the approximation length. We wanted to give them a tool that could recreate the movements and changes in the data and allow them to vary the visualized approximation length in real time. We accomplished this goal by parallelizing the bulk of the application: reading, interpolating, and processing the particle positions.

Interpolating each point between two time-differentiated values and two approximation-differentiated values is extremely resource intensive. Interpolating these four values on the CPU simply would be too taxing due to the limited parallel capacity. However, since each particle can be processed independently of the rest of the data, the problem is *embarrassingly parallel*. We can therefore offload a significant amount of the workload onto the GPU. Sending the data into video memory as textures, we use vertex shaders to read and interpolate the four variations of each particle. After determining the position and velocity of a particle, the shader can compute the offset-adjusted position and the color. As a result of this approach, the users can smoothly interpolate across both time and approximation length in real time while rendering millions of particles.

### 4.2 Findings

Continuing to use the aforementioned spatial and parallel coordinates views, scientists can see how the data's correlations and uncertainties vary over time (see figure 6 and figure 8). Figure 6 displays the particles at a very early time (upper panel) and in their final state (lower panel). Initially the particles are distributed very smoothly following a Gaussian random field, and under the influence of gravity, they end up in tight structures. The velocity and location uncertainty, which represented variation among

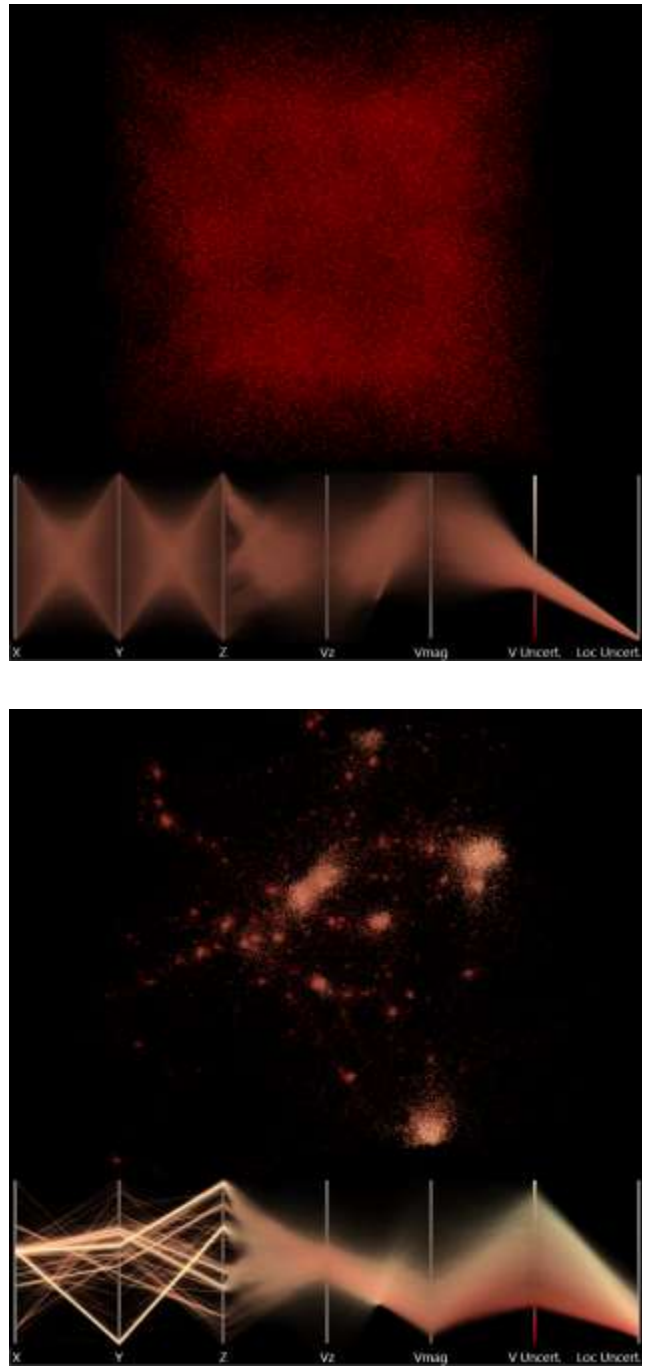

Figure 6. The **top** image shows time step 50. The particles have high velocities and are still fairly even distributed. The parallel coordinates view provides a lot of insight about the data. The pattern between the X, Y, and Z axes on the left and the even brightness show that the data is evenly distributed and has little or no clusters. The velocity uncertainty and location uncertainty (rightmost axes) have log scales and show the limited variation at this time step caused by the approximation (which the dim red color also illustrates).

The **bottom** image shows time step 0. The vast majority of the particles have coalesced into clusters with high uncertainty. The bright lines between the spatial axes on the left represent those dense uncertain clusters. Both uncertainty axes show a significant increase compared with the top image.

simulator code results, now represent variation caused by different approximation lengths. The scientists wanted to examine many aspects of the particles, so they can choose to correlate the color of the particles based on those properties. Coloring by velocity uncertainty and animating across time, the user can easily spot any periods of rapid variation growth or even possible reduction (which scientists hope to find). We also provide the user with the option of drawing a line between extreme positions, so color can be used for another property such as velocity magnitude (see figure 8).

One important statistic that is of interest to the scientists is the mass function. The mass function measures the number of bound structures, so-called halos (clusters are the largest halos), in a certain mass bin. In [19] it was found that the number of halos and therefore the mass function was different if the simulation was started at different epochs.

One question that arises is why fewer halos formed when the simulation was started late. Figure 7 provides a very important clue to this question. The upper left panel shows the earlier start and two halos (a large one and a smaller one) are pointed out. These two structures are much fuzzier in the right panel (late start). Since the halo detecting algorithm is based on finding nearest neighbors, particles in the fuzzy outer part of a halo will not be identified as belonging to that halo. Therefore, the mass of the halos (being the sum of all particles belonging to the group) in the right panel will be lower, and small structures will be missed.

The mass function will therefore be lower overall if the simulation is started too late.

The next question to answer is how early the simulation must begin to capture all halos. In order to answer this question, convergence studies have to be carried out. If the simulation is started early enough, the results should be the same when compared with the results of an even earlier start. Figure 8 shows an example of how visualization can aid the convergence studies. The simulation began at three different times, and the panels show variations of the particles at different time steps from two different starting points. At very early times, the difference between the two starts is very small, indicating that the start at 150 might have been sufficiently early. As the simulation progresses, the differences become larger. Starting from the fourth panel, nonlinearities amplify the differences in the structures, and the image accordingly shows much more variation. The large scale structures start to form at this point and the uncertainty lines appear as a shell around where the cluster will eventually form. In ongoing work, higher order approximation schemes for the initial conditions are investigated, and the current tool will be very helpful in exploring the differences between the different schemes.

## 5 CONCLUSION

The accuracy of simulation data is of crucial importance to cosmologists. Ongoing and upcoming surveys will map out the structures in the universe to very high precision. In order to interpret these observations, simulations have to be as accurate as the observations. Many aspects of research are based on these simulation results, and understanding the complex spatial and temporal patterns of the accuracy levels is difficult to accomplish using only quantitative analysis. By allowing the data to be interactively visualized, scientists can have a better understanding of where and to what extent simulation variation and approximation may impact results. Scientists commonly measure simple statistics from the simulations, such as mass functions or two-point correlation functions. Differences in these statistics due to different simulation algorithms, starting redshifts, particle loadings, force resolution, and so on have to be understood at the percent level accuracy. Visualizing the differences in the simulation outputs under different code settings can be very helpful in gaining a better understanding of their causes. At the accuracy level required, the obtained insight can be pivotal for progress in understanding the error properties of complex codes.

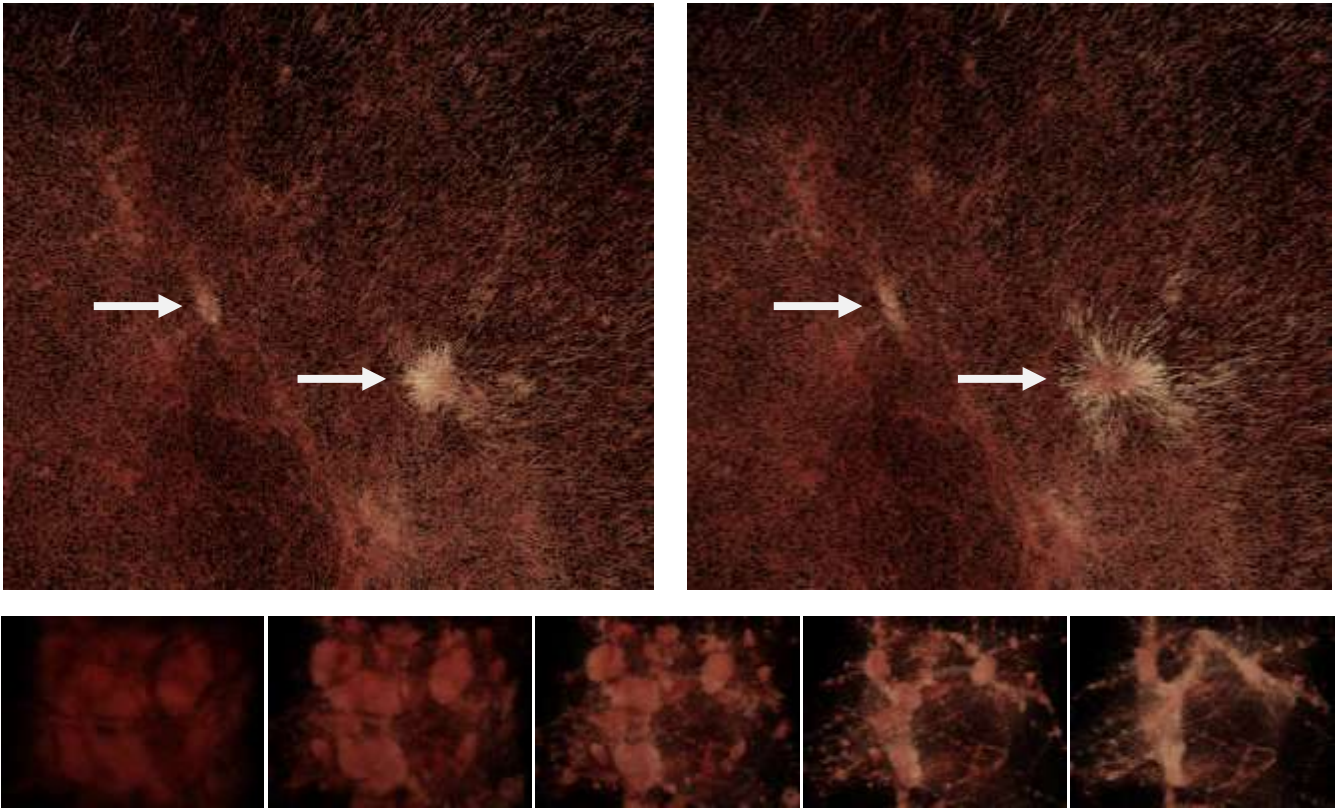

Figure 7. The **top** images are zoomed into a region of particles at redshift 10. Each particle is rendered as a small line oriented and sized by the velocity vector, and the color is representative of velocity uncertainty. The left image shows the particle positions and velocities if the Zel'dovich approximation were taken to redshift 250. The particles in the right image were approximated to redshift 50. The structural formation and coalescence of the particles is much more significant with the reduced approximation (left) than with the longer approximation (right), and varied density of the bright clusters best demonstrates this observation.
The **bottom** images show the dataset as it progresses from redshift 50 to the present (redshift 0). Each particle is rendered as a small point, and color represents velocity.

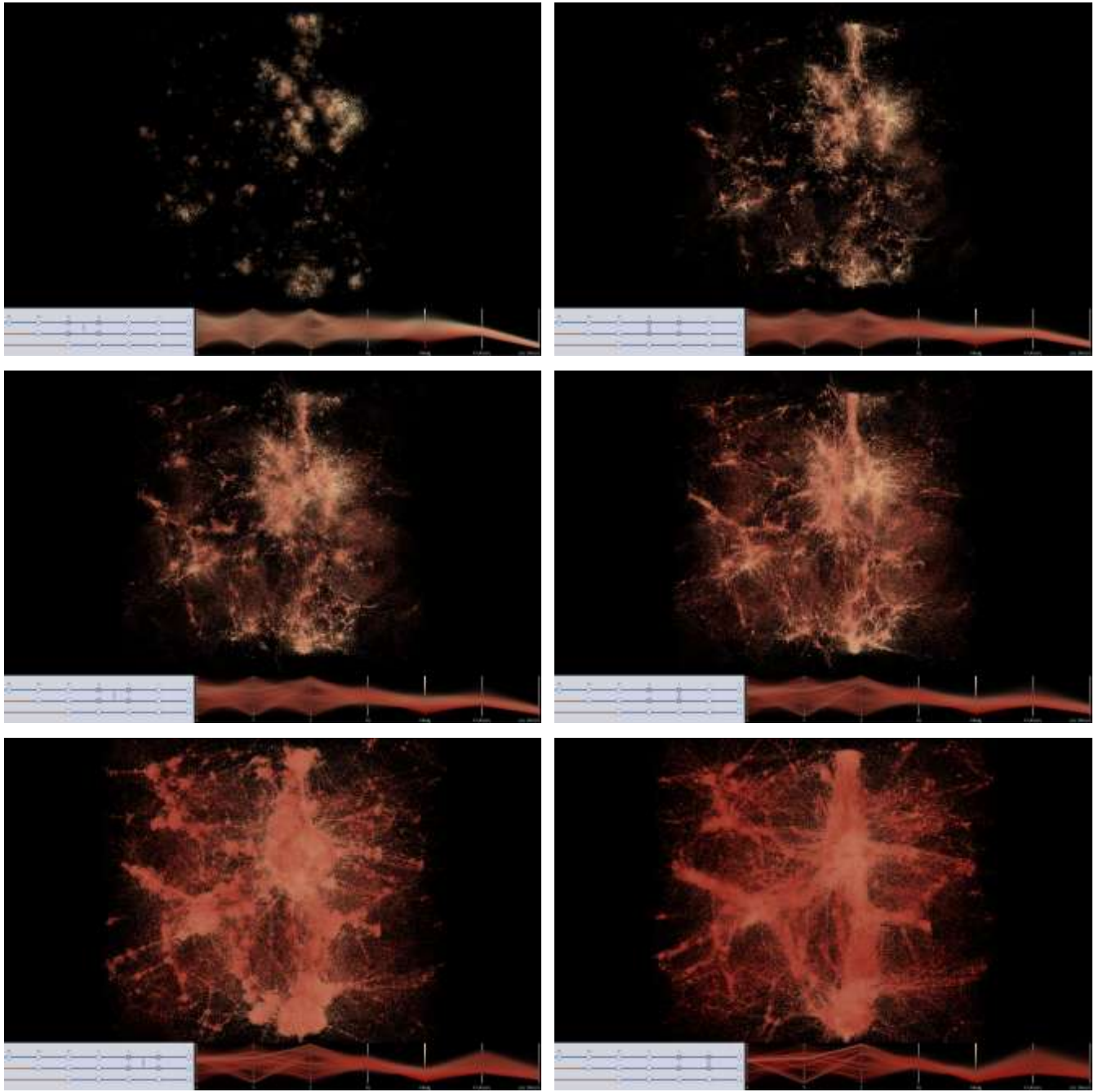

Figure 8. These images show a progression from redshift 50 to redshift 1. For each particle, a line colored by velocity extends from the position in one series to the position in another. As time advances, the variation between the series grows and becomes more structured. The time and series selection interface can be seen in the bottom left of each image with the darkened circles representing the currently active timesteps.

A parallel coordinate control interface is helpful in visualizing correlations in particle data. While established as a powerful technique for many visualization systems, we use a parallel coordinates view as an interface tool to extend the limitations of a spatial view. This interface is helping scientists explore and gain insight into uncertain data that is too complex for a spatial view to display alone. As scientists continue to generate ever larger sets of uncertain, multidimensional, spatial data, they need a means of selecting the type of information to examine. By using modern programmable graphics hardware, we can give scientists a level of detail and interactivity and a range of options that were impossible only one year ago. The new performance benefits also allow animation, once only relegated to pre-rendered clips, to be an integral, insight-enhancing, and perceptually beneficial feature. This work provides an example application that can view complex, multidimensional, and time-variant data as well as critical uncertainty information. Presenting the uncertainties as axes that are no different from standard variables, we improve the simplicity and flexibility of the users' tasks. A benefit of this system is that interacting with uncertainty in the same manner as other variables can provide insight into correlations of the uncertainty without eliminating the ability to visualize other variables.

## 6 ACKNOWLEDGEMENT

This research was supported in part by the U.S. National Science Foundation through grants CCF-0634913, IIS-0552334, CNS-0551727, and OCI-0325934, and the U.S. Department of Energy through the SciDAC program with Agreement No. DE-FC02-06ER25777. Thanks to the Cosmic Data ArXiv for making the datasets publicly available.